\documentclass[%
aip,
rsi,%
amsmath,amssymb,
preprint,%
floatfix 
]{revtex4-1}
\usepackage{graphicx}
\usepackage{dcolumn}
\usepackage{threeparttable}
\usepackage{gensymb}
\usepackage{inputenc}  
\usepackage{bm} 
\usepackage{amsmath}  
\usepackage{amsfonts} 
\begin{document}
\title{Nanoantenna enhanced radiative and anisotropic decay rates in  monolayer-quantum dots}
\author{Laxmi Narayan Tripathi}
\affiliation{ Department of Physics, Indian Institute of Science, Bangalore, India}
\email{ltripathi@physik.uni-wuerzburg.de}
\author{M.Praveena}
\affiliation{ Department of Physics, Indian Institute of Science, Bangalore, India}
\author{Ben Johns}
\affiliation{ Department of Physics, Indian Institute of Science, Bangalore, India}
\author{Jaydeep Kumar Basu}
\affiliation{ Department of Physics, Indian Institute of Science, Bangalore, India}

\begin{abstract}
Nanoantenna enhanced ultrafast emission from colloidal quantum dots as quantum emitters is required  for fast quantum communications. On chip integration of such devices require a scalable and high throughput technology. We report self-assembly lithography technique of preparing hybrid of gold-nanorods antenna over a compact CdSe quantum dot monolayer. We demonstrate resonant and off resonant gold nanorod antenna enhanced radiative and anisotropic decay. Extensive simulations explain the mechanism of the decay rates and the role of antenna in both random and a compact monolayer of quantum dots. The study   could  find applications in quantum dots display and quantum communications devices.  
\end{abstract}
\maketitle
\section{Introduction} 
 Colloidal quantum dots (QDs) are known to be single photon emitters \cite{ Hoang2016} and are desperately needed for quantum communications, quantum information processing and quantum computers etc. Optical antenna \cite{Giannini2011b,Novotny2011} have been shown to enhance the spontaneous emission \cite{Eggleston2015} as well as ultrafast emission \cite{Hoang2016}  through radiative decay rate enhancement of the  photons for fast quantum communication.
  Resonant excitation of optical antennas such as metal nanopartics/rods enhance the electromagnetic field dramatically via localized surface plasmon resonance     \cite{Novotny2011,Novotny2006,Bharadwaj2009}.  A single QD can experience  this strong electromagnetic environment via coupling to this surface plasmon modes provided QDs lie within the penetration depth of electric field  of surface plasmon  \cite{Neogi2005}. The decay rate of the QDs can be significantly modified by the resonant excitation of QDs and metal nanoantenna \cite{Haridas2010c,Russell2012}. Fabrication of aligned and isolated optical antenna over mono-layer QDs is an important step in realization of such optical interaction. We need a  high throughput and scalable technology to fabricate preferably aligned optical antenna over a compact monolayer of QDs. 
     
   Here, we provide  optimum physical parameters to transfer such antenna over monolayer of QDs.  We combine a directed self-assembly  technology using Langmuir-Blodgett (LB) \cite{Tripathi2013,Haridas2011a,Tripathi2014} and controlled dip coating,  a  high throughput and scalable method.  A controlled dip coating  provides a better scheme  in transferring  aligned GNRs. In the dip coating process, a known concentration, C (mg/ml) of CTAB capped GNRs in water and a KSV mini trough (Finland) dipper were used to insert a QD film inside  the GNR solutions,  waited for  dip time, $ \tau $ and then taken out with  speed, u.  We have optimized the condition for transfer of aligned GNRs from water. The parameters like dipping time ($ \tau $), dipping speed (u) and area fraction,  $ \phi $ is given in the Table \ref{tab1}.  The dipping method imparts an overall directionality to the GNR, along the dipping direction while the density of  GNRs were controlled  by the dipping speed, density of GNRs solutions.  Recently, we have also shown long range emission enhancement  \cite{Tripathi2014} due to resonant optical antenna from a mono-layer of QDs. Here,  we demonstrate radiative decay rates enhancements  from a mono layer of QDs due to isolated and aligned optical antenna. We have performed extensive finite difference time domain (FDTD) calculations for  decay rates enhancements of a dipole emitter as a function of separation from  both resonant and non resonant antenna.
   
   To prepare a monolayer of CdSe quantum  dots (QDs), we synthesized   CdSe QDs of  mean  core diameter of 10 nm  (Fig.~1(b)) with emission wavelength, 650 nm, using  method  developed   by Peng et al  \cite{Peng2001}. Photoluminescence (PL) spectrum for the QDs is shown in Fig. 1 (a). Gold nanorods (GNRs) of  aspect ratios (AR) 2 and 3, hereafter, referred as AR2 and AR3 respectively were  chemically synthesized using methods as described earlier \cite{Sau2004,Tripathi2014}. The shape and size of GNR  were estimated using transmission electron microscope (TEM) micrograph. The concentrations of  GNR in  water was determined from   Inductively coupled plasma optical emission spectroscopy (ICP-OES)  (Perkin-Elmer). The UV visible absorption data were used to estimate the concentration of GNRs in a diluted solution. The absorption spectra of thin film of GNRs with AR2 and AR3 are shown in the Fig. 1(a). We can see that the longitudinal surface plasmon  resonance (LSPR) peak of AR 2  is overlapping with emission peak of CdSe QDs therefore, we call AR2 GNRs as resonant antenna. Similarly LSPR peak of AR3 GNR is not overlapping with PL of QDs hence we call it off-resonant GNRs antenna.
    
      \begin{figure}
      	\centering
      	\includegraphics[scale=0.6]{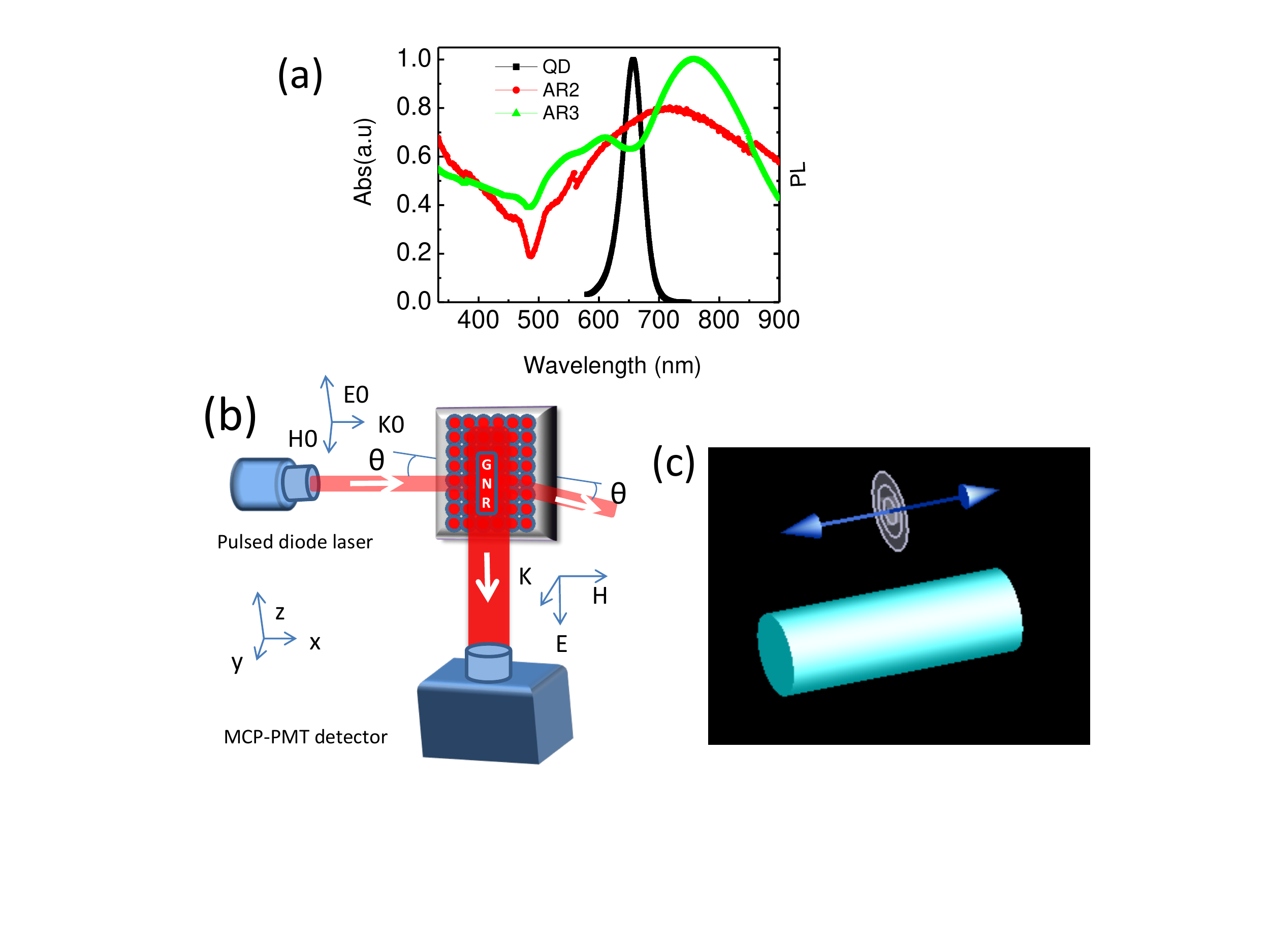}
      	\caption{ (a) Absorption spectra  of dense film of GNR antenna, both AR2; AR3 and PL spectrum of CdSe QDs in toluene. (b) TCSPC set up: pulsed 633 nm diode laser and showing the  \textbf{p} polarization of excitation and emission and Micro-channel plate- photomultiplier (MCP-PMT) detector. (c)  Schematic diagram used for simulating  a GNR antenna and a dipole emitter with a polarization state parallel to the length of the antenna.}
      \end{figure}

 \begin{table}
   \begin{threeparttable}
 \caption { \label{tab1} \emph{
 \textbf{Optimum parameters for transfer of aligned  GNRs on  a CdSe QDs monolayer.}} }
 \tabcolsep=0.3 cm
 \begin{tabular}{lcccccc}
 	 & \cr
 \hline
 Sample\tnote{1} & C \tnote{1} & $ \tau $ \tnote{2} & u \tnote{3} & $\phi$ \tnote{4} &  \cr
  index & mg/ml & min & mm/min & \% &  \cr
  \hline
 $S_{1}$ & 0.003 & 1 & 10 & 1.04 & \cr
 
 $S_{2}$ & 0.003 & 15 & 10 & 2.64 &  \cr
 
 $S_{3}$ & 0.003 & 30 & 10 & 3.94 &  \cr
 
 $S_{4}$ & 0.006 & 1 & 10 & 1.63 &  \cr
 
 $S_{5}$ & 0.006 & 15 & 10 & 2.51 &  \cr
 
 $S_{6}$ & 0.006 & 30 & 10 & 6.25 &  \cr
 
 $S_{7}$ & 0.012 & 1 & 10 & 5.89 &  \cr
 
 $S_{8}$ & 0.012 & 15 & 10 & 7.10 &  \cr
 
 $S_{9}$ & 0.012 & 30 & 10 & 9.21 &  \cr
 \hline
 \end{tabular}
 
 \begin{tablenotes}
     \item[1] Concentration of  CTAB capped GNRs in mg/ml in water.
     \item[2] Time duration which the substrate  was inside water.
     \item[3] Speed of the dipper with respect to water.
     \item[4] Fraction of area covered by GNRs on the  QD monolayer was calculated  from TEM images using Image J software.
  \end{tablenotes}
  \end{threeparttable}
 \end{table}
    
    We  transferred a  compact  LB monolayer of CdSe QDs and used it as substrate to transfer an aligned layer of GNR by dipping it for one minute inside a GNR solution. Then it was pulled out in a controlled way using a  dipper at the rate of 10 mm/min.  To transfer low  number density  of GNR over the film,  concentration of GNRs in the solution was   intentionally   kept low ($ \equiv $ 0.01 mg /ml). The film was then  dried in vacuum for 12 hrs. Transfer and orientation of rods along the dipping direction  was confirmed by AFM. The typical ratio of GNR : QD was  found to be 1:800. Table 1 shows different set of physical parameters for aligned and isolated GNR antenna transfer.  Let us define hybrid of GNR with AR2 as A1 and hybrid of GNR with AR3 is A2.  We  performed time correlated single photon counting (TCSPC)  measurements using Horiba Scientific Fluoro cube-01-NL.  Pulsed picoseconds  laser  diode  with  pulse duration 70 ps and  repetition rate of 1 MHz was used as excitation source. A peak preset of 20,000 counts was used during data collection. 
    \begin{figure}
    	\centering
    	\includegraphics[scale=0.7]{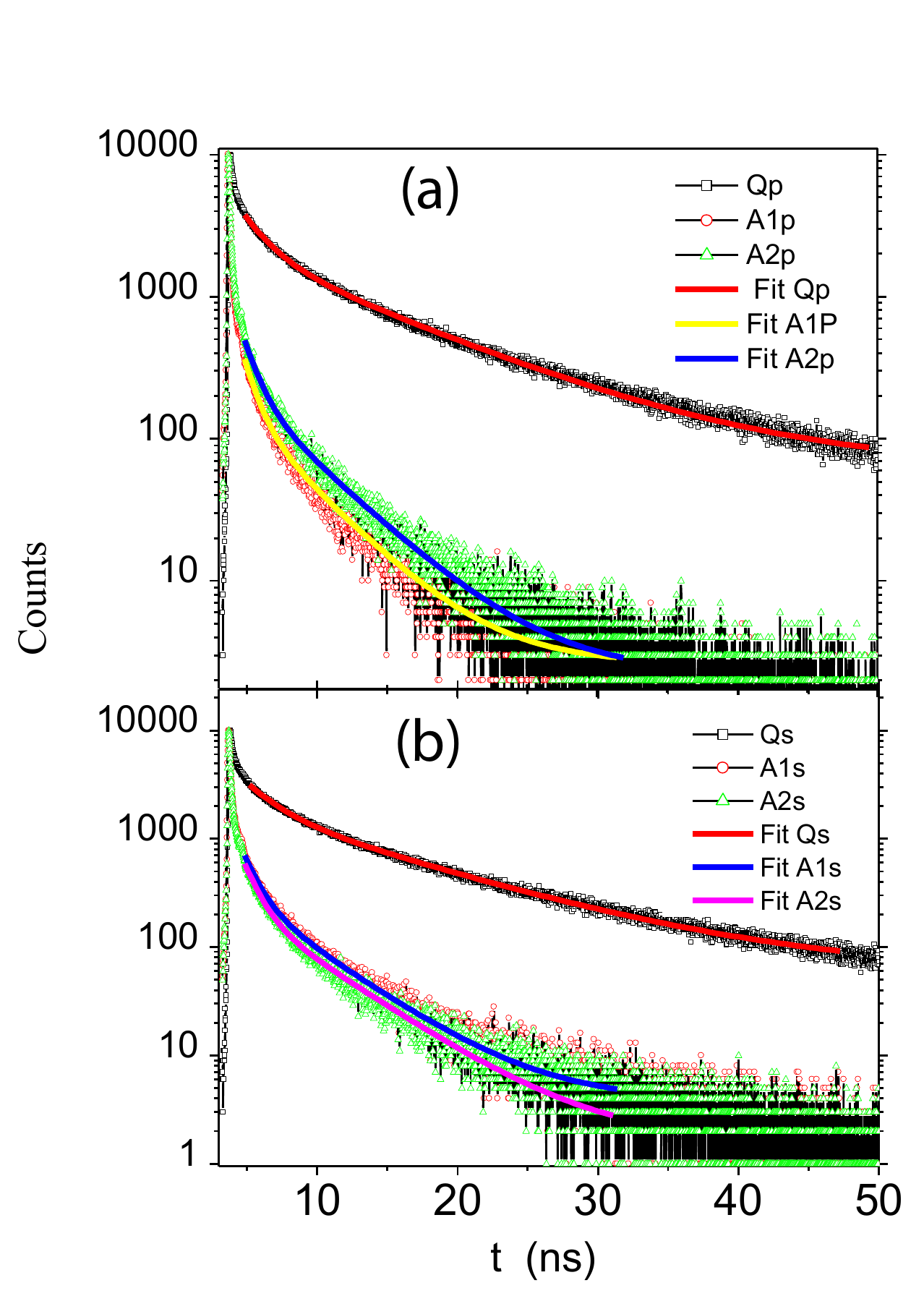}
    	\caption{\label{decay_sol}  Log-linear plot of  TCSPC data on \textit{solution sample} under (a) \textbf{p} excitation for QDs in chloroform (Qp), hybrid A1 (A1p); hybrid A2 (A2p). (b) \textbf{s} excitation for QDs (Qs), hybrid A1 (A1s); hybrid A2 (A2s). The exponential fit is shown by the thick lines.}
    \end{figure}
    
      A schematic diagram of  the TCSPC set up  is shown in Fig.~1(c). A polarized laser of 630 nm was incident with a grazing angle , $ \sim 5 \degree $ with  film sample (Fig. 1(c)). The grazing angle ensure the large area excitation of the hybrid, therefore increases the total  PL intensity and  reduces the data collection time. To compare the exciton dynamics of the QDs monolayer  with isolated and random  QDs we also collected time resolved photoluminescence spectra of 1 : 10000 ratio of GNR and  QDs hybrid  in chloroform  after mixing  1 mg/ml of CdSe QDs  to 0.1 mg  Octadecane thiol capped GNRs. Fig. 2 shows the time resolved  PL counts collected from the dilute hybrid solution of QDs and GNR optical antenna in chlorofom in a closed cuvette.The data were collected using \textbf{p} polarized,
       along z axis  as shown in the Fig. 1(c) and \textbf{s} (along y axis as shown in the Table Fig. 1(c)). Here, sample lies in XZ plane. The \textbf{p} excitation at 630 nm excites LSPR of AR2 antenna.
 To extract the life time of excitons in the hybrid of  QDs and GNR antenna for both in film and solution, we fitted the decay curve using mostly double  or three exponential  equation if the fitting is not good using the equation $ I=I_{0}+a_{1}e^ \frac{-(t-t_{0})} {\tau_{1}}+a_{2}e^ \frac{-(t-t_{0})} {\tau_{2}}+a_{3}e^ \frac{-(t-t_{0})} {\tau_{3}} $  where $ t_{0}$ , $I_{0}$ are offset (shift) values in time (t) and intensity (I) in the data;  $a_{1}, a_{2}, a_{3}$  are pre-exponential factors,   $\tau{1}, \tau{2}, \tau{3}$ are three components of  exciton life time. Mean life time of excitons ($\tau$),  for example,  for double exponential decay   was calculated   as ( $\tau = \frac{a1\tau^2+a1\tau^2}{a1\tau+a2\tau}$) \cite{Lakowicz2006}. The total decay  rate,  $ \Gamma $ for a QD in excited state from the hybrid monolayer can be written  in terms of radiative decay rate , $\Gamma_{R}$  and non-radiative decay rate $\Gamma_{NR} $ as  	$ \Gamma = \Gamma_{R}+\Gamma_{NR}  $ and  $ \Gamma =\frac{1}{\tau } $. Knowing the   quantum yield (QY) (10\%) of the QDs we calculated the radiative decay rate using the relation : QY = $ \frac{\Gamma_{R}}{\Gamma} $. Here, we measured the QY in toluene and assumed that it is same for QDs transferred onto the substrates. This is justified assumptions as QY is calculated per unit of QD.

  From Fig. 2 we can see that the  PL decay faster for  the hybrid of resonant GNRs (AR2)  than the bare QDs solutions under both
  \textbf{s} and \textbf{p} type of excitation. This is similar to what we observed on film data as shown in Table.~\ref{filmfitpar_tab}. Although the difference between  the mean lifetimes of the QDs in solution under \textbf{p} and \textbf{s}  excitation,  in each hybrid (A1 or A2),  is not significantly large yet it is non zero. This might be due to  random motion of GNRs and QDs in the solution.  We have also calculated anisotropy in life time  defined as, $ G_\tau  = \left| \frac{\tau_{p}-\tau_{s}}{\tau_{p}+\tau_s}\right| $ where $\tau_{p}$ and $\tau_{s}$  are  the average PL life time of CdSe QDs in the hybrid assemblies under \textbf{ p } and \textbf{s} excitation using 630 nm laser respectively. Further, from the Table .~\ref{filmfitpar_tab} we can see that the anisotropy induced in the A1 hybrid for the solution is 3 fold larger than the bare CdSe QDs solution. This suggests that anisotropy in life time is introduced due to  interaction of  plasmonic field with QDs.
     
       \begin{figure}
         	\centering
         	\includegraphics[scale=0.7]{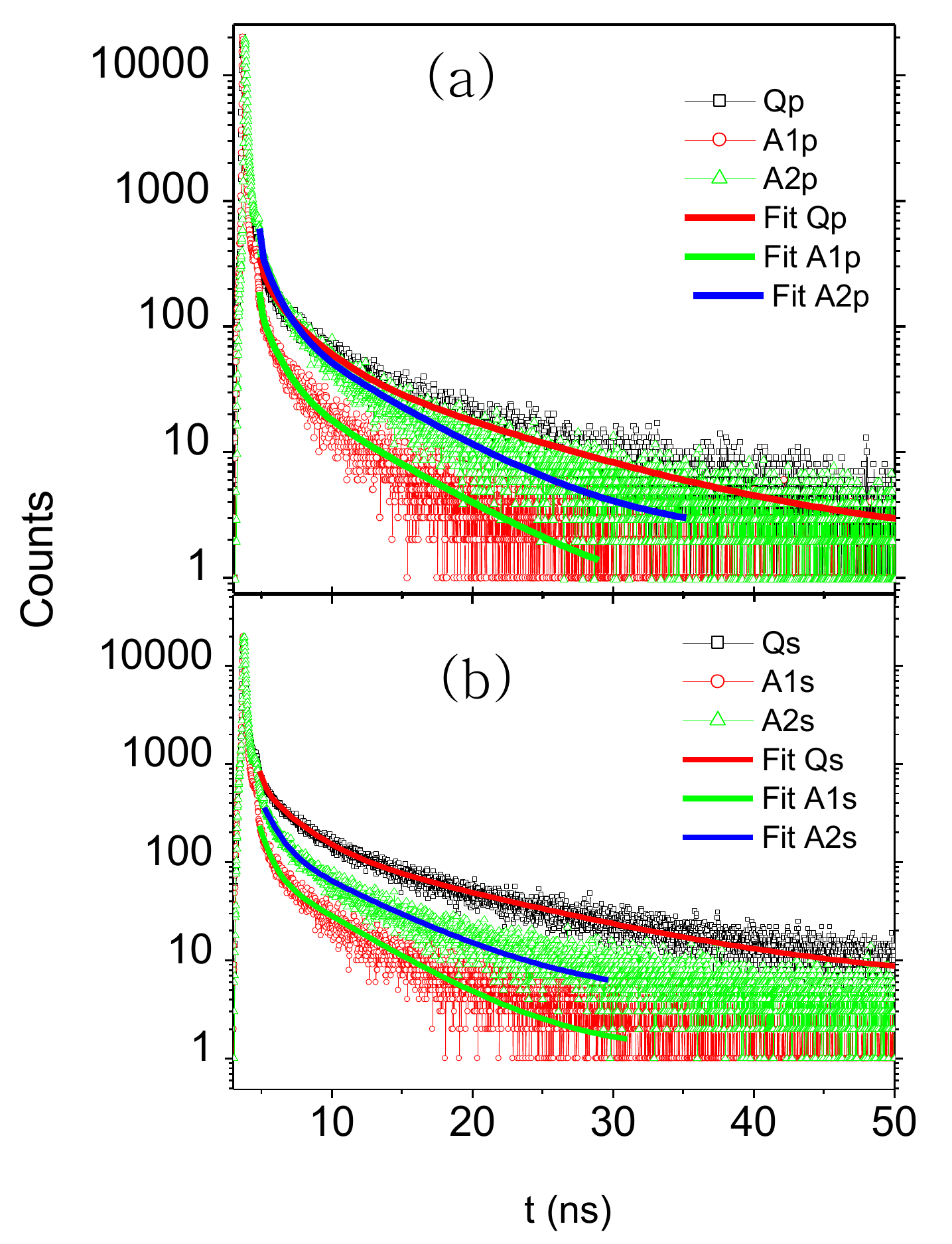}
         	\caption{  \label{decay_antenna}  Log-linear plot of TCSPC data on \textit{film sample} under. (a) \textbf{p} excitation for QDs in chloroform (Qp), Hybrid A1 (A1p); Hybrid A2 (A2p). (b) \textbf{s } excitation for QDs (Qs), Hybrid A1 (A1s); hybrid A2 (A2s). The exponential fit is shown by the thick lines. }
         \end{figure}
        
 To obtain further insight into the mechanism of light matter interaction in hybrid film we have performed  TCSPC measurement on these samples. As shown in the Table.~\ref{filmfitpar_tab},  reductions of the decay life times in the sample, A1  and A2 is obvious. Further, we observed  a  two  fold enhancement of the radiative decay rate. Again we observed clear spectral effect in the lifetime anisotropy for A1 samples (resonant excitation). Despite the fact that lifetime measurements have been done with a broad beam and  hence the over all orientation alignment of GNR within the illuminated area being much smaller  it is clear that the samples (A1) shows  higher lifetime anisotropy than the intrinsic anisotropy of the QDs monolayer  as shown in Table \ref{filmfitpar_tab}. However, we believe the anisotropy with 633 nm excitation  would be stronger if a much smaller illuminated  area  and high density aligned nanorods antenna are used. 
 
 \begin{table}
 	\centering
 	\caption { \label{filmfitpar_tab} Average life time $\tau_{film} $ and $\tau_{sol}$;  radiative decay rate $\Gamma_{R-film}$ and $\Gamma_{R-sol}$;  anisotropy in life time $G_{\tau-film }$ and $G_{\tau-sol }$;  for  QDs (Q)  and hybrid A1 and hybrid A2 film and solution respectively for each polarization, P (\textbf{p} and \textbf{s}).}
 	\tabcolsep=0.1cm
 	\begin{tabular}{lccccccccccccccccccc}
 		 & \cr
 		\hline
 Sample & P & $\tau_{fim} $ & $\tau_{sol} $ & $\Gamma_{R-Film}$ & $\Gamma_{R-Sol}$  & $G_{\tau-film}$  & $G_{\tau-Sol }$& \cr 
         &  & ns  & ns& $ns^{-1}$  & $ns^{-1}$     &          &       &           &\cr \hline
 		$Q$  & \textbf{p}& 7.58  & 9.08  & 0.0132  & 0.0110   & 0.04  &   0.02    & \cr
 		$A1$ & -         & 4.65  & 3.39  & 0.0215  & 0.0295   & 0.07  &   0.06    & \cr
 		$A2$ & -         & 4.45  & 3.80  & 0.0225  & 0.0263   & 0.04  &   0.02      &\cr
 		$Q$  & \textbf{s}& 8.15  & 9.50  & 0.0123  & 0.0105   &       &           & \cr
 		$A1$ & -         & 4.05  & 3.83  & 0.0247  & 0.0261   &       &           &   \cr
 		$A2$ & -         & 4.85  & 3.91  & 0.2062  & 0.0256   &       &           & \cr
 		\hline
 	\end{tabular}
 \end{table}

     Finite Difference Time Domain (FDTD) simulations (Lumerical solutions) were used  to estimate the decay rates for dipoles radiating in the presence of a GNR.  The decay rates of the dipole in the presence of resonant (AR1) and off resonant GNR (AR2) were calculated by measuring the power emitted from the dipole at 660 nm wavelength. The GNR is approximated by a gold cylinder (Fig. 1 (d)), and the optical  data for gold is taken  from Johnson and Christy \cite{JOHNSON1972}. For simulating the lifetime in solution, an averaging over different spatial orientations were done – namely axial, parallel and normal orientations of  a dipole with respect to the nanorod. In case of film, for \textbf{p}-polarized excitation, an average of axial and parallel orientations  and for \textbf{s}-polarized excitation, the normal orientation was simulated. The size of the simulation region was $ 1\times 1 ~\mu m^2 $ , with a uniform mesh of 2 nm over the GNR and  a dipole. A broadband plane wave source of intensity 1 ($ \frac{W}{m^2} $) is used in all the electric field simulations. Electric field monitors are used to record local fields around the rod. The dipole moment of the source is fixed at $ 2.8*10^{-31} $ C-m for all simulations and the wavelength is taken as the emission wavelength of the QD. The electric field modification at the excitation frequency   of the QD in the vicinity of a  resonant GNR is shown in Fig. (4e-f). We see a strong  E field enhancement (22) near  GNR antenna and  which decay to zero  within few nanometer away.  For a dipole transition, the decay rate in an inhomogeneous environment is  related to the classical power output of the dipole in the same environment by the relation $\frac{\Gamma}{\Gamma_{0}}=\frac{P} {P_{0}} $ \cite{Novotny2006} where $ \Gamma $ is a radiative decay rate, P, $ P_{0} $ are  emission powers of dipole emitters in the presence of the antenna  and in the free space respectively. We obtain the normalized total power of the dipole in the presence of the GNR ($ \frac{P}{P_{0}} $), which gives us the normalized total decay rate. The total power emitted by the dipoles is also calculated using power monitors arranged as a box around the dipole. This can be written as the sum of radiated and non-radiated power, $ P = P_r + P_{nr} $. The rate of transfer (non radiative power) is calculated as the difference of the total power and the radiated power.  Normalizing these quantities with the free space values gives the ratios of radiative and non radiative decay rates.  $\frac{\Gamma_{r}}{\Gamma_{0}}=\frac{P_r}{P_{0}} $  and $\frac{\Gamma_{nr}}{\Gamma_{0}}=\frac{P_{nr}}{P_{0}} $.
 \begin{figure}
 	\centering
 	\includegraphics[scale=0.8]{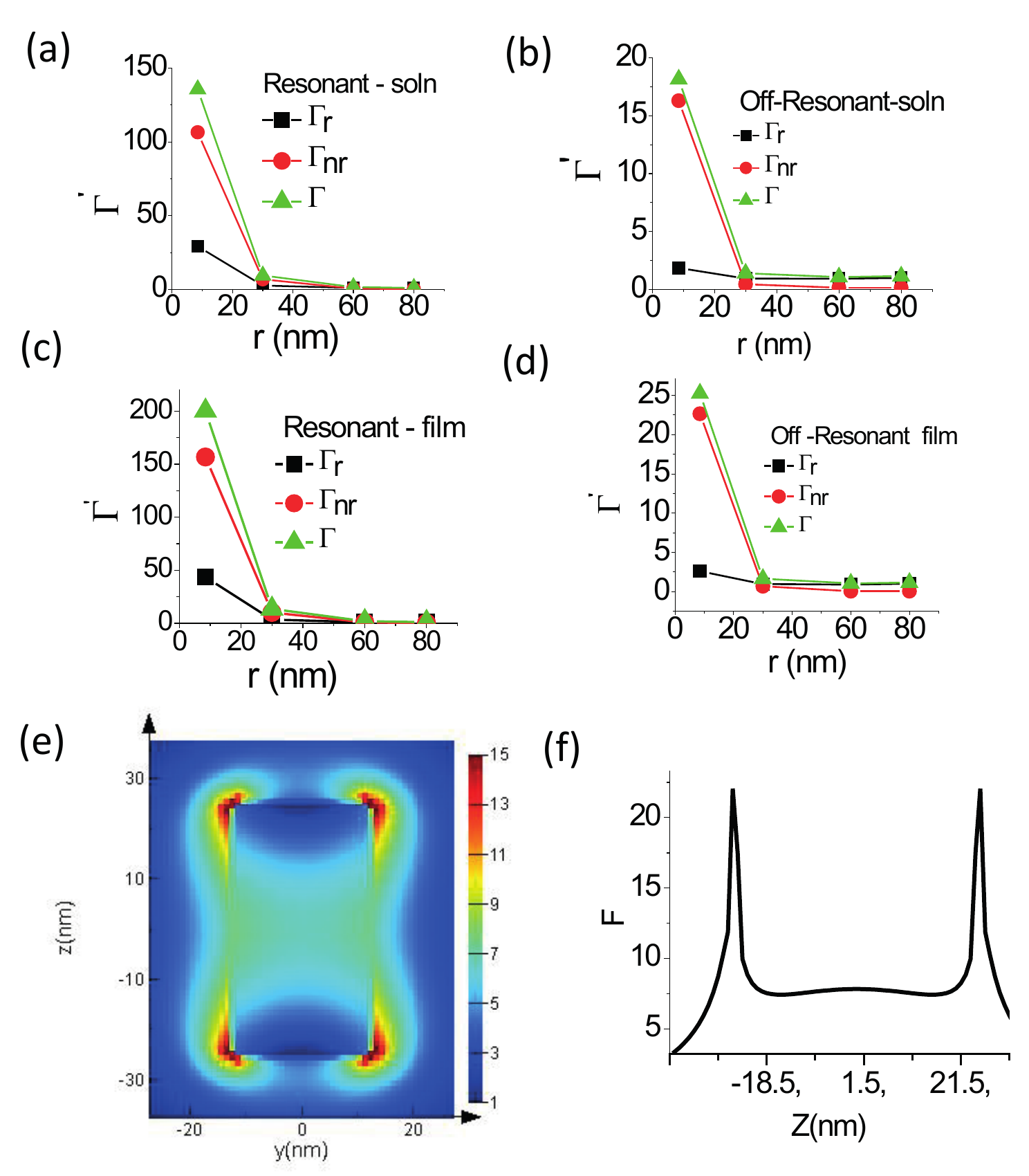}
 	\caption{\label{FDTD} Calculated  decay rates enhancements ($ \Gamma^{'} $) for a dipole in solution and film  under \textbf{p} excitation  as a function of separation between the dipole and (a) GNR antenna AR2 in Chloroform   (b) GNR antenna AR3 in chloroform  (c) GNR antenna AR2 in film  (d) GNR antenna AR3 in film. (e) Electric field enhancement map around a GNR AR2 under \textbf{p} excitation. (f) Line profile of electric field enhancement (F) values extracted from (e) along the length of the antenna.}
 \end{figure}
 
 From the Fig. 4 (a-d) it is seen that very close to the metal particle, the non radiative decay dominates the radiative process. The radiative rate enhancement for a dipole emitter near a metal nanoparticle is mainly due to coupling with the dipole mode of the surface plasmon, whereas the emitter can couple with all modes of the surface plasmon to transfer energy. At very close distances, the coupling with the higher modes becomes stronger leading to very high non radiative decay. At larger distances, the coupling with higher modes becomes weak and the radiative processes dominate in this regime.  FDTD results show that the modification in decay rates is greater for the resonant system (Fig.~4 (a),~4(c)) as compared to the off resonant (Fig. 4(b), 4(d)) system. In solution, the orientations are not fixed and are expected to be random.  To know the mean separation between QDs and GNR, we  first calculated the mean distance between the QDs, $ r_{mean} $ using the probability distribution of  randomly moving QDs and GNRs in chloroform as $ r_{mean}  = 0.55396 n ^{-1/3}$  \cite{Chandrasekhar1943}  where n is number of QDs per unit volume. Using the known concentration (1 mg/ml) of QDs and the mean radius as 5 nm ,  $ r_{mean} $ turns out to be 80.6 nm. Considering length of  GNR as 50 (60) nm in case of AR2 (AR3) the minimum  surface to surface separation is 10.3 (5.3) nm.  So we have calculated the decay rate enhancement  as  a function of separation (r) between the dipole emitter and GNRs antenna (Fig. 4 (a-d)). The orientation-averaged radiative decay rate enhancement in the Hybrid A1 at the separation of 30 nm  is found to be 2.6 which is closer to experimental radiative decay rate of 2.7 (Table. \ref{filmfitpar_tab}). In comparison, the  off resont hybrid at 30 nm separation in solution  does not show any decay rate enhancement ($ \Gamma '  \approxeq 1$). For the close packed film, the resonant system has a mean radiative decay rate enhancement of 3.5 (Fig. 4(c)) at 30 nm separation  for p polarization as opposed to experimental radiative decay rate. ($ \approxeq 2 $), whereas the decay rates for \textbf{s} polarization (not shown here) is almost unchanged. The off resonant cases show very little deviation from the bare QD system.  For the resonant system, the  decay rates are more enhanced in the case of \textbf{p} polarized emission as compared to the \textbf{s} polarized case, indicating the role of the longitudinal surface plasmon resonance in modifying the local density of states of the emitter, thereby increasing its emission properties.
 
 In conclusion, in this letter,  we provide optimum  physical parameters to transfer aligned  and low density GNR antenna over monolayer of QDs and report  resonant and off resonant GNR antenna induced radiative and anisotropic  decay from the compact monolayer and isolated  QDs in solution. We observed enhanced decay rates when the quantum emitters are  in the vicinity of the resonant optical antenna. Further an anisotropic decay was introduced  both in the isolated QDs and in the monolayer film by virtue of interaction of the optical antenna.  Our  fundamental study  could be very useful for the scientific community in QDs display and  quantum communications devices etc.
      
     We acknowledge Department of Science and  Technology (Nanomission), India, for financial support and   Advanced Facility for Microscopy and Microanalysis, Indian Institute of Science, Bangalore, for access to TEM  measurements. M. Praveena acknowledges UGC, India, forc
     financial support

\end{document}